\documentclass[referee]{raa}            

\usepackage[dvipsnames]{xcolor}
\usepackage{ulem}
\usepackage{graphicx,times}             
\usepackage{natbib}
\usepackage[T1]{fontenc}
\usepackage{amssymb,amsmath}

\usepackage{graphics}
\usepackage{bm}

\usepackage{multicol}
\usepackage{multirow}
\usepackage{pdflscape}
\usepackage{lscape}
\usepackage[figuresright]{rotating}

\usepackage{chngpage}
\usepackage{array}
\usepackage{color}
\usepackage{enumerate}
\usepackage{algorithm}
\usepackage{algorithmic}
\usepackage{indentfirst}
\usepackage[bottom]{footmisc}
\bibpunct{(}{)}{;}{a}{}{,}
\usepackage[pass,a4paper]{geometry}
\usepackage[a4paper=true,pagebackref=true]{hyperref}
\hypersetup{colorlinks = true, linkcolor = green, anchorcolor = red, citecolor = blue, filecolor = red, pagecolor = red, urlcolor = red}
\newcommand\micron{\mbox{$\mu$m}}%

\newcommand{\beq}{\begin{equation}}
\newcommand{\eeq}{\end{equation}}

\newcommand{\mstar}{\rm{\bf M_{\bf \star}}}

\newcommand{\mhi}{\rm \bf{M_{HI}}}

\begin{document}

  \title{The H{\sc i} gas fraction scaling relation of the Green Pea galaxies
}

   \volnopage{Vol.0 (20xx) No.0, 000--000}      
   \setcounter{page}{1}          

   \author{Siqi Liu 
      \inst{1,2}
      \and A-Li Luo
      \inst{1,2,3}
      \and
          Wei Zhang
          \inst{1,2}
          \and
          Yan-Xia Zhang
          \inst{1,2}
          \and
          Xiao Kong
          \inst{1}
          \and 
          Yong-Heng Zhao 
          \inst{1}
   }
   \institute{CAS Key Laboratory of Optical Astronomy, National Astronomical Observatories, Beijing 100101, China; {\it lal@nao.cas.cn,xtwfn@bao.ac.cn}\\
        \and
             University of Chinese Academy of Sciences, Beijing 100049, China\\
        \and 
        College of Computer and Information Management \& Institute for Astronomical Science, Dezhou University, Dezhou 253023, China\\
\vs\no
   {\small Received 2022 Nov 15; accepted 2023 Mar 16}}

\abstract{
Green Pea galaxies are compact galaxies with high star formation rates.  However, limited samples of Green Pea galaxies have H{\sc i} 21 cm measurements. 
   Whether the H{\sc i} gas fraction ($f_{\rm H{\sc i}} \equiv \rm M_{H \sc I}/M_{\star}$) of Green Pea galaxies follows the existing scaling relations between the $f_{\rm H{\sc i}}$ and NUV-$r$ color or linear combinations of color and other physical quantities needs checking. 
    Using archival data of H{\sc i} 21cm observations, we investigate the scaling relation of the NUV-$r$ color with the $\rm M_{H \sc I}/M_{\star}$ of 38 Green Pea galaxies, including 17 detections and 21 non-detections.  
   The H{\sc i} to stellar mass ratios ($f_{\rm H{\sc i}}$) of Green Pea galaxies deviate from the polynomial form, where a higher H{\sc i} gas fraction is predicted given the current NUV-$r$ color, even with the emission lines removed.
   The blue sources (NUV-$r$<1) from the comparison sample (ALFALFA-SDSS) follow a similar trend.
    The H{\sc i} gas fraction scaling relations with linear combination forms of $\rm -0.34(NUV-r) - 0.64 \log(\mu_{\star,z}) + 5.94 $ and $\rm -0.77 \log \mu_{\star,i} + 0.26 \log SFR/M_{\star}+8.53$, better predict the H{\sc i} gas fraction of the Green Pea galaxies. 
    In order to obtain accurate linear combined forms, higher-resolution photometry from space-based telescopes is needed.
\keywords{galaxies: general -- galaxies: starburst -- radio lines: galaxies}
}

   \authorrunning{S. Liu, A-L. Luo, W. Zhang \it{et. al. } }            
   \titlerunning{Green Pea galaxies H{\sc i} gas fraction scaling relation}  

   \maketitle

\section{Introduction}
Green Pea galaxies are well known for their unique color, compactness, high star formation rate (SFR), and are located in isolated environments \citep{2009MNRAS.399.1191C}.
They are local analogs to the high-$z$ Ly$\alpha$ emitters due to their low metallicity, low dust content, and high ionization \citep{2011ApJ...728..161I,2016ApJ...820..130Y,2019ApJ...872..145J}.

The observations of the H{\sc i} 21cm provide information about the neutral atomic mass, which is the primary fuel for star formation in these galaxies.  
Presently, there are insufficient studies \citep{2014ApJ...794..101P,2019ApJ...874...52M,2021ApJ...913L..15K} with limited samples of the Green Pea galaxies.
Lyman Alpha Reference Sample (LARS) from \citet{2014ApJ...794..101P} provides an upper limit of one Green Pea galaxy (LARS 14) based on the non-detection of the H{\sc i} 21 cm emission line from the Green Bank Telescope (GBT). 
\citet{2019ApJ...874...52M} observed one Green Pea galaxy with the Very Large Array (VLA) and found no 21 cm hydrogen hyperfine-structure line detected.  
\citet{2021ApJ...913L..15K} provide 19 detections and 21 upper limits on the Green Pea galaxies at $0.02 < z < 0.1$ from the Arecibo Telescope (Arecibo) and the GBT.  
They find the H{\sc i}-to-stellar mass ratio ($f_{H \sc I}$) in Green Pea galaxies is consistent with the star-forming galaxies in the local Universe but with a much shorter gas depletion time of about 0.6 Gyr.  

Scaling relations provide a cheap estimator for the H{\sc i} gas fractions in the galaxies. 
However, there is a lack of blue samples to verify different sets of scaling relations. 
In this work, we use all of the currently available archival H{\sc i} 21 cm measurements of the Green Pea galaxies to verify the H{\sc i} gas fraction scaling relations. 
We find that the scaling relation of the NUV-$r$ with H{\sc i} gas fraction for these Green Pea galaxies deviates from the relation in \citet{2021A&A...648A..25Z} of local star-forming galaxies, where a lower fraction of the H{\sc i} gas is observed given the current NUV-$r$ color, even with the emission lines removed.
 The linear combination gas fraction scaling relations \citep{2009MNRAS.397.1243Z,2010MNRAS.403..683C,2012MNRAS.424.1471L,2021A&A...648A..25Z} combining the surface mass density, surface brightness, with color or specific star formation rate (sSFR) better predict the H{\sc i} gas fraction of the Green Pea galaxies, especially with the form of $\rm -0.34(NUV-r) - 0.64 \log(\mu_{\star,z}) + 5.94 $ and $\rm -0.77 \log \mu_{\star,i} + 0.26 \log SFR/M_{\star}+8.53$. 

The structure of the paper is as follows.  
In Section \ref{sec:data}, we present H{\sc i} 21 cm observation of the Green Pea galaxies and the comparison sample of local star-forming galaxies from ALFALFA-SDSS \citep{2020AJ....160..271D}. 
In Section \ref{sec:H I}, we discuss the NUV - $r$ color scaling relation, other factors that impact the gas fraction estimation, and the linear combination estimators of the gas fraction.
In Section \ref{sec:result}, we summarize the result and propose future work.

In this work, we use Wilkinson Microwave Anisotropy Probe (WMAP) 9 cosmology \citep{2013ApJS..208...19H}, 
AB magnitudes \citep{1983ApJ...266..713O},
a Chabrier initial mass function (IMF; \citet{2003PASP..115..763C}), and \citet{2003MNRAS.344.1000B} stellar population synthesis (SPS) models.

\section{Data}
\label{sec:data}
\subsection{Green Pea galaxies with H{\sc i} 21 cm observations }
\subsubsection{H{\sc i} 21 cm observations}
There are limited Green Pea galaxies with H{\sc i} observations, including \citet{2014ApJ...794..101P,2019ApJ...874...52M} and \citet{2021ApJ...913L..15K}. 
All of the samples in this work are from \citet{2021ApJ...913L..15K}.

We include one Green Pea galaxy Lyman Alpha Reference Sample (LARS) 14 from \citet{2014ApJ...794..101P} according to SDSS and \textit{HST} images.
GBT observation shows no H{\sc i} 21 cm emission line. Thus an upper limit of the H{\sc i} mass is given with the 3$\sigma$ detection.
\citet{2019ApJ...874...52M} observed one Green Pea galaxy with the VLA and found no 21 cm hydrogen hyperfine-structure line detected.  
The upper limit of the H{\sc i} mass is obtained assuming a 3$\sigma$ detection threshold $S_{\rm 21cm} = 0.78$ mJy and characteristic emission line width of $W_{\rm 21cm} = 36.1$ km s$^{-1}$.  
However, this source is duplicated with GP1608+3528 from \citet{2021ApJ...913L..15K} observed with Arecibo, with non-detection of H{\sc i} 21 cm emission line from both observations.
For consistency, we keep the non-detection result from Arecibo for the following discussions.

With a deep search of the H{\sc i} 21 cm emission line based on Arecibo and GBT of the Green Pea galaxies, \citet{2021ApJ...913L..15K} have obtained 19 detections and 21 upper limits of the H{\sc i} mass. 
They are pre-selected from the spectroscopically identified Green Pea galaxies \citep{2019ApJ...872..145J} based on the $B$-band luminosity \citep{2014MNRAS.444..667D} to have detectable H{\sc i} 21 cm.  
The selected sample covers a wide range of absolute $B-$band magnitudes ($-20.0 \le M_B \le -16.1$), and gas-phase metallicity ($\rm -7.6 \le 12+(O/H) \le 8.35$), which is also statistically consistent with the parent \citet{2019ApJ...872..145J} sample in metallicity, stellar mass, and absolute $B-$band magnitude. 
Both the Arecibo and GBT observations use the L-wide receiver and two orthogonal polarizations. 
The Arecibo observations use a 25 MHz band sub-divided into 4096 spectral channels, while the GBT observations use a 23.44 MHz bandwidth subdivided into 8192 channels. 
The total integration time on each source ranges from $0.75-4.5$ hrs.
All the data are analyzed following the standard procedures with {\sc gbtidl}, and each spectrum is checked for RFI and systematic effects.
The H{\sc i} 21 cm emission lines are detected at $\ge 5\sigma$ level. 
For the non-detections, the upper limits are estimated assuming that the emission line profile is Gaussian with a full width at half maximum (FWHM) of 50 km s $^{-1}$.

We find the positions for the sources in SDSS DR16 \citep{2020ApJS..249....3A} given the name and the redshift of the sources and obtain the \texttt{cModelMag} for the $ugriz$ bands for each of these compact galaxies.  
For GP0844+0226, we cannot find its SDSS spectrum; and thus exclude it for further analysis.
We only include the sources located within the GALEX footprint, which results in 17 H{\sc i} 21-cm emission line detections with NUV photometry and 21 H{\sc i} 21-cm emission line non-detections with NUV photometry. 

We also apply the $k$-correction with the code \citet{2010MNRAS.405.1409C} and \citet{2012MNRAS.419.1727C}. 
Because these Green Pea galaxies are isolated sources, we have visually checked them individually in the SDSS and GALEX images to avoid source mismatching.  
We inherit the results of H{\sc i} masses from \citet{2021ApJ...913L..15K}, which are calculated with $\int S dV$ from the H{\sc i} 21-cm emission profiles, according to \citet{1994ARA&A..32..115R}.
For the extinction correction, the Galactic extinction from the dust maps of \citet{1998ApJ...500..525S} has been applied to both the SDSS magnitudes and the GALEX magnitudes. 
Furthermore, to be noted, as illustrated in Fig. 6 in \citet{2021A&A...648A..25Z}, internal extinction will not affect this H{\sc i} gas fraction with the NUV-$r$ relation. 

\begin{figure}[!ht]
  \centering
  \includegraphics[width=0.7\hsize]{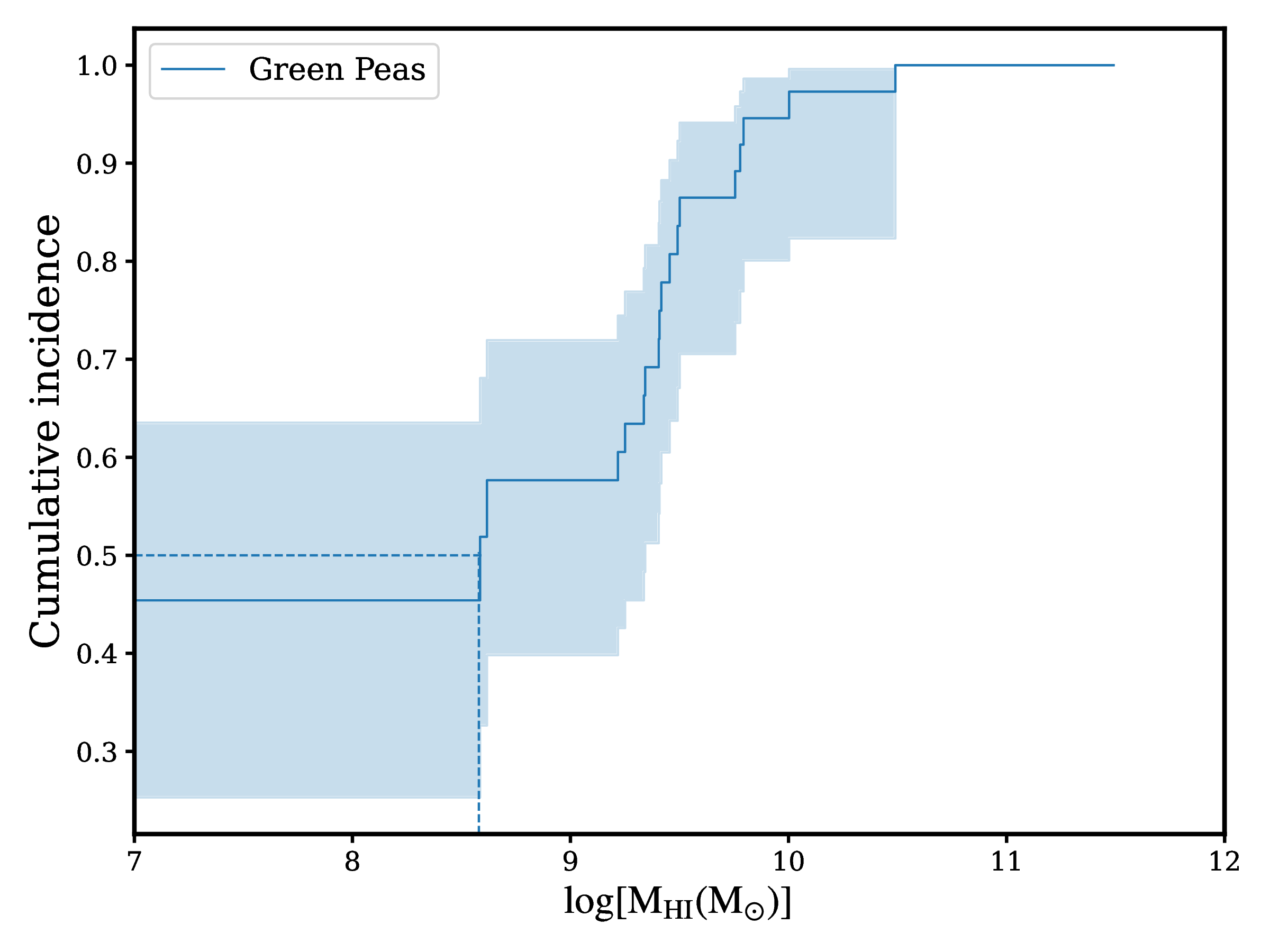}
      \caption{
     Cumulative incidence curve of the uni-variate $\rm M_{H \sc I}$ distribution estimated with Kaplan-Meier estimator with confidence bands.
     The dashed lines show that the probability of detection is 50\% at $\rm \log (M_{H \sc I}/M_{\odot})=8.58$. 
 }
         \label{fig:cif_GP}
  \end{figure}

The uni-variate distribution of $\rm M_{H \sc I}$ can be estimated with the Kaplan-Meier \citep{doi:10.1080/01621459.1958.10501452} estimator with confidence bands. 
We demonstrate the cumulative incidence distribution for the $\rm M_{H \sc I}$ of the Green Pea galaxies in Fig.\ref{fig:cif_GP} generated with \texttt{lifelines}.  
At $\rm \log (M_{H \sc I}/M_{\odot})=8.58$, the probability that the H{\sc i} 21 cm emission can be detected is 50\%.

\subsubsection{Mass and SFR}
Instead of using the given stellar mass measurement and the SFR measurement from \citet{2021ApJ...913L..15K}, we re-calculate these two quantities to obtain more accurate measurements for the analysis in this work.
We cross-match the sources with the AllWISE Multiepoch Photometry Table  \citep{2010AJ....140.1868W,Wright_2019-os} to obtain the $W_1$ to $W_4$ photometry.

For the stellar mass measurement, we fit the sources with multiwavelength photometry with CIGALE \citep{2005MNRAS.360.1413B,2009MNRAS.397.1243Z,2019A&A...622A.103B} combining the GALEX NUV, SDSS $ugriz$, and AllWISE $W_1$ to $W_4$ photometry.  
For the configuration of the fitting, we use the delayed$-\tau$ star formation history, BC03 \citep{2003MNRAS.344.1000B}, Chabrier IMF \citep{2003PASP..115..763C}, nebular emission lines, the dust attenuated modified starburst model, the dust emission model from \citet{2012MNRAS.425.3094C}, and the \citet{2006MNRAS.366..767F} AGN model.

For the SFR measurement, we combine the IR and UV photometry using the method in \citet{2020AJ....160..271D} Section 3.2, where the SFR tracer is robust against extinction.
The corrected NUV spectral energy density is $\nu L_{\nu,\rm corrected} = \nu L_{\nu,\rm NUV} + 2.26 \nu L_{\nu} (22~\micron)$.
The SFR based on the corrected NUV \citep{2011ApJ...741..124H,2012ARA&A..50..531K,2020AJ....160..271D} is calculated as 
\begin{eqnarray}
\rm \log_{10} (SFR_{NUV_{corr}}) = \log_{10}(\nu L_{\nu,corrected})-43.17,
\end{eqnarray}
For the source without $W_4$ detection (GP\_id J0808+1728), we use the measured result from \citet{2021ApJ...913L..15K}.

\begin{figure}[!ht]
  \centering
  \includegraphics[width=0.7\hsize]{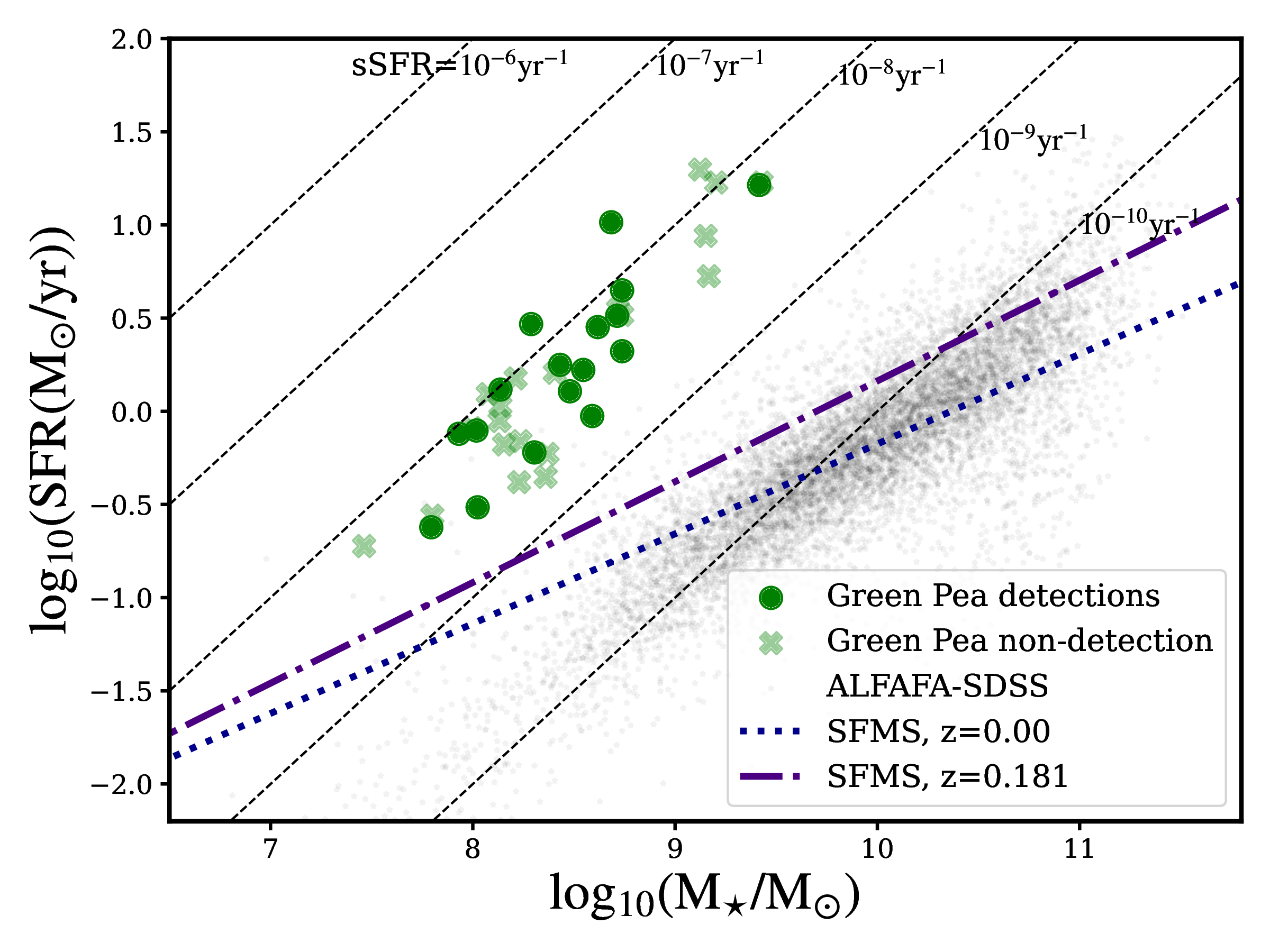}
      \caption{Stellar mass vs. SFR for the Green Pea galaxies, where the green circles demonstrate the sources with H{\sc i} 21 cm detections and the green crosses mark the non-detections.  
      The ALFALFA-SDSS samples \citep{2020AJ....160..271D} as the comparison samples are plotted with gray dots in the background.
      We also mark the SFMS relation \citep{2014ApJS..214...15S} at $z=0.0$ (dotted line) and $z=0.181$ (dash-dot line).  
      The Green Pea galaxies have higher SFRs compared with the SFMS.}
         \label{fig:mass-sfr}
  \end{figure}

We compare the mass-SFR relation for these samples with the star-forming main sequence (SFMS) \citep{2014ApJS..214...15S} in Fig.\ref{fig:mass-sfr}. 
The main-sequence SFR-$M_{\star}$ relation from \cite{2014ApJS..214...15S} is as follows:
\begin{eqnarray}
    \log \rm SFR (M_{\ast}, t) &=& (0.84 \pm 0.02 - 0.026 \pm 0.003 \times t )\log M_{\ast} \nonumber\\
   & &\quad - (6.51 \pm 0.24 -0.11 \pm 0.03 \times t),
\end{eqnarray}
where $t$ is the age of the universe in Gyr. 
The Green Pea galaxies have much higher SFR compared with the SFMS from $z=0.0$ to $z=0.181$, where the sSFRs range from $10^{-9} \rm yr^{-1}$ to $10^{-7} \rm yr^{-1}$. 

We list the H{\sc i} 21 cm observation results, the mass and SFR measurements from this work in Table \ref{table:GP}.
\par
\begin{table*}[]
\begin{center}
\caption[]{Green Pea galaxies with H{\sc i} 21 cm measurements}\label{table:GP}
\resizebox{\textwidth}{!}{%
\begin{tabular}{lllllllllll}
\hline
\hline
\textbf{Green Pea} &$\textbf z$ & \textbf{Tel.$^{1}$} &\textbf{$\bf \nu_{\rm \bf 21\,cm}$} &\textbf{ $\bf \int \bf S {\rm \bf d}\bf V$ }& \textbf{$\mhi$} & \textbf{SFR}$^{2}$ & \textbf{$\mstar$}$^{3}$ & \textbf{NUV-$\bf r$$^{4}$}  &\textbf{$\bf R_{50,z}$$^{5}$}  &\textbf{$\bf R_{50,\it HST}$$^{6}$}\\
\textbf{identifier} &    &      &  \textbf{MHz}      & \textbf{Jy~km~s$^{-1}$} & \textbf{$ 10^8 \; \textrm{M}_{\odot}$ } & \textbf{$\textrm{M}_{\odot}$~yr$^{-1}$}  & \textbf{{$10^8 \; \textrm{M}_{\odot}$}}      &   & \textbf{arcsec}  & \textbf{arcsec}
\\ \hline
J0007+0226 & $0.0636$ & $1$ & $ 1335.52 $ & $ < 0.062$         & $ < 12 $         & $0.8859$ & $1.355 $ &   $0.65 \pm 0.22$ &  0.6240 &  --\\
J0036+0052 & $0.0282$ & $1$ & $ 1381.42 $ & $0.585 \pm 0.034$  & $22.0 \pm 1.3$   & $0.2385$ & $  0.6232 $    & $0.499 \pm 0.024$ &  1.033 &   --\\
J0159+0751 & $0.0611$ & $1$ & $ 1338.65 $ & $ < 0.038$         & $ < 6.9$         & $0.8123 $ & $  0.9915 $  &   $0.02 \pm 0.29$  &  2.042 &   --\\
J0213+0056 & $0.0399$ & $2$ & $ 1365.88 $ & $0.419 \pm 0.022$  & $31.7 \pm 1.7$   & $0.7889 $ & $ 1.042$   &   $-0.772 \pm 0.057$ &  0.7293 &   --\\
J0801+3823 & $0.0376$ & $2$ & $ 1368.89 $ & $ < 0.051$         & $ < 3.4$         & $0.4453$ & $ 2.288 $  &  $0.981 \pm 0.025$ &  0.9185 &   --\\
J0808+1728 & $0.0442$ & $1$ & $ 1360.28 $ & $ < 0.044$         & $ < 4.1$         & $0.45^{7}$ & $   0.2902$  &  $-0.13 \pm 0.09$ &  0.9449 &   --\\
J0844+0226 & $0.0911$ & $1$ & $ 1301.81 $ & $0.065 \pm 0.013$  & $26.1 \pm 5.4$   & $16.39$ & $ 25.93 $    &  $0.751 \pm 0.086$ &  1.110 &   --\\
J0852+1216 & $0.0759$ & $1$ & $ 1320.19 $ & $ < 0.050$         & $ < 14 $         & $19.82$ & $13.28$  &  $0.129 \pm 0.032$ &  1.207 &   --\\
J0926+4427 $^{8}$ & $0.1807$ & $2$ & $1202.71$ & $<1.7$ &$<3100$ & $16.90$ &$15.97$  & $0.393 \pm 0.022$ &  0.7927 &   --\\
J0942+4110 & $0.0460$ & $2$ & $ 1357.97 $ & $ < 0.125$         & $ < 13$          & $1.502 $ & $1.627$  &  $0.867 \pm 0.038$ &  2.587 &   --\\
J1015+3054 & $0.0918$ & $1$ & $ 1301.04 $ & $ < 0.037$         & $ < 15 $         & $8.735 $ & $14.17$  & $0.774 \pm 0.030$  &   0.8601 &   --\\
J1024+0524 & $0.0332$ & $1$ & $ 1374.76 $ & $0.074 \pm 0.014$  & $3.85 \pm 0.73$  & $1.775 $ & $2.700$   & $0.2419 \pm 0.0083$  &  1.683 &   --\\
J1108+2238 & $0.0238$ & $1$ & $ 1387.37 $ & $0.155 \pm 0.016$  & $4.14 \pm 0.43$  & $0.3054$ & $1.056$   &  $0.758 \pm 0.051$ &  1.248 &  0.9640\\
J1134+5006 & $0.0260$ & $2$ & $ 1384.44 $ & $0.799 \pm 0.065$  & $25.4 \pm 2.1$   & $1.668$ & $ 3.516$    &  $0.2260 \pm 0.0091$ &  1.057 &   --\\
J1148+2546 & $0.0451$ & $1$ & $ 1359.07 $ & $3.182 \pm 0.078$  & $309.0 \pm 7.6$  & $4.458 $ & $ 5.470 $   &  $-0.163 \pm 0.036$ &  1.262 &   --\\
J1200+2719 & $0.0819$ & $1$ & $ 1312.91 $ & $0.310 \pm 0.028$  & $100.6 \pm 9.0$  & $3.256$ & $ 5.169$   &  $-0.188 \pm 0.083$ &  0.8862 &   --\\
J1224+0105 & $0.0398$ & $2$ & $ 1365.99 $ & $ < 0.063$         & $ < 4.8$         & $0.5906$ & $2.341 $  &  $0.632 \pm 0.025$  &  1.242 &   --\\
J1224+3724 & $0.0404$ & $2$ & $ 1365.25 $ & $ < 0.077$         & $ < 5.9$         & $1.069$ & $ 1.369 $  & $0.023 \pm 0.058$ &   0.9575 &   --\\
J1226+0415 & $0.0942$ & $1$ & $ 1298.10 $ & $ < 0.073$         & $ < 32 $         & $3.528 $ & $ 5.218$  &  $0.069 \pm 0.037$ &  0.9608 &   --\\
J1253-0312 & $0.0227$ & $2$ & $ 1388.89 $ & $0.235 \pm 0.021$  & $56.9 \pm 5.1$   & $10.34$ & $  4.832  $  &  $0.569 \pm 0.021$ &  1.088 &   --\\
J1319+0050 & $0.0477$ & $1$ & $ 1355.78 $ & $0.234 \pm 0.025$  & $21.7 \pm 2.3$   & $0.9432 $ & $ 3.892$  & $1.175 \pm 0.031$ &  0.8985 &   --\\
J1329+1700 & $0.0942$ & $1$ & $ 1298.16 $ & $ < 0.086$         & $ < 37 $         & $16.89 $ & $ 2.679$  &  $0.169 \pm 0.042$ &  0.7646 &   --\\
J1345+0442 & $0.0304$ & $1$ & $ 1378.47 $ & $0.650 \pm 0.026$  & $28.5 \pm 1.1$   & $1.308 $ & $ 1.369 $  &  $-0.538 \pm 0.029$   &  0.8460 &  0.4181 \\
J1359+5726 & $0.0338$ & $2$ & $ 1373.93 $ & $ < 0.095$         & $ < 5.1$         & $3.282 $ & $ 5.480$  &  $-0.011 \pm 0.020$ &  1.034 &   --\\
J1423+2257 & $0.0328$ & $1$ & $ 1375.24 $ & $ < 0.025$         & $ < 1.3$         & $0.6637$ & $1.425 $  &   $0.389 \pm 0.061$ &  1.024 &   --\\
J1432+5152 & $0.0256$ & $2$ & $ 1384.94 $ & $ < 0.079$         & $ < 2.4$         & $0.4168$ & $1.695 $  &   $0.819 \pm 0.018$ &  1.169 &   --\\
J1448-0110 & $0.0274$ & $2$ & $ 1382.50 $ & $ < 0.054$         & $ < 1.9$         & $1.609$ & $2.542$  &  $0.885 \pm 0.012$ &  1.059 &   --\\
J1451-0056 & $0.0432$ & $2$ & $ 1361.59 $ & $0.288 \pm 0.029$  & $25.6 \pm 2.6$   & $1.280$ & $3.020 $  &  $-0.152 \pm 0.015$ &  0.9594 &   --\\
J1455+3808 & $0.0277$ & $2$ & $ 1382.13 $ & $0.855 \pm 0.047$  & $31.0 \pm 1.7$   & $0.7575$ & $ 0.853$  &  $0.023 \pm 0.029$ &  0.9091 &  0.5770 \\
J1509+3731 & $0.0326$ & $2$ & $ 1375.58 $ & $< 0.088$          & $< 4.4$          & $1.239$ & $1.354 $  & $0.882 \pm 0.063$ &  1.617 &   --\\
J1509+4543 & $0.0481$ & $2$ & $ 1355.18 $ & $0.543 \pm 0.081$  & $60.0 \pm 8.9$   & $2.101$ & $ 5.471$   & $0.308 \pm 0.040$ &  1.160 &   --\\
J1518+1955 & $0.0751$ & $1$ & $ 1321.19 $ & $ < 0.041$         & $ < 11 $         & $5.309$ & $ 14.69$  &  $0.509 \pm 0.019$ &  0.9274 &   --\\
J1545+0858 & $0.0377$ & $1$ & $ 1368.76 $ & $0.302 \pm 0.019$  & $17.8 \pm 1.1$   & $2.938$ & $ 1.942$  &  $0.052 \pm 0.045 $ &  0.6910 &   --\\
J1547+2203 & $0.0314$ & $1$ & $ 1377.15 $ & $ < 0.053$         & $ < 2.5$         & $0.6915$ & $1.721$  & $1.436 \pm 0.025$ &   1.121 &   --\\
J1608+3528 $^{9}$& $0.0327$ & $1$ & $ 1375.38 $ & $ < 0.12$          & $< 0.61$         & $0.1891$ & $0.2809$  &  $0.68 \pm 0.16 $ &  1.022 &   --\\
J1624-0022 & $0.0313$ & $1$ & $ 1377.27 $ & $0.134 \pm 0.021$  & $62.2 \pm 9.8$   & $2.832 $ & $4.472 $   &   $0.729 \pm 0.039$ &  1.253 &   --\\
J2114-0036 & $0.0447$ & $2$ & $ 1359.59 $ & $0.173 \pm 0.020$  & $16.5 \pm 1.9$   & $0.6017$ & $ 2.014 $  &  $0.297 \pm 0.084$ &  1.187 &   --\\
J2302+0049 & $0.0331$ & $1$ & $ 1374.91 $ & $ < 0.050$         & $ < 2.6$         & $0.2756$ & $  0.6311 $  & $0.102 \pm 0.026$ &  0.9076 &   --\\
\hline
\hline
\end{tabular}
}
\end{center}
\footnotesize{$^1$ Following the notation in \citet{2021ApJ...913L..15K}, we use Arecibo~$\equiv 1$, GBT~$\equiv 2$ to note the telescope used for the observations.}\\
\footnotesize{$^2$ The SFR is measured combining UV and IR photometry \citep{2020AJ....160..271D}}\\
\footnotesize{$^3$ The stellar mass is measured with multi-wavelength photometry fitting combining the GALEX NUV, SDSS $ugriz$, and AllWISE $W_1$ to $W_4$ photometry.}\\
\footnotesize{$^4$ The NUV-$r$ color has been $k$-corrected.}\\
\footnotesize{$^5$ From SDSS DR16 \citep{2020ApJS..249....3A}.}\\
\footnotesize{$^6$  The entries are from \textit{HST} COS instrument \citep{2019hst..prop16054K}.}\\
\footnotesize{$^7$ Inherited from \citet{2021ApJ...913L..15K}.}\\
\footnotesize{$^8$ From \citet{2014ApJ...794..101P}.}\\
\footnotesize{$^9$ In duplication with the VLA observation of \citet{2019ApJ...874...52M}.}
\end{table*}

\subsection{Star forming galaxies with H{\sc i} 21-cm observation}

\begin{figure}[!ht]
  \centering
  \includegraphics[width=0.7\hsize]{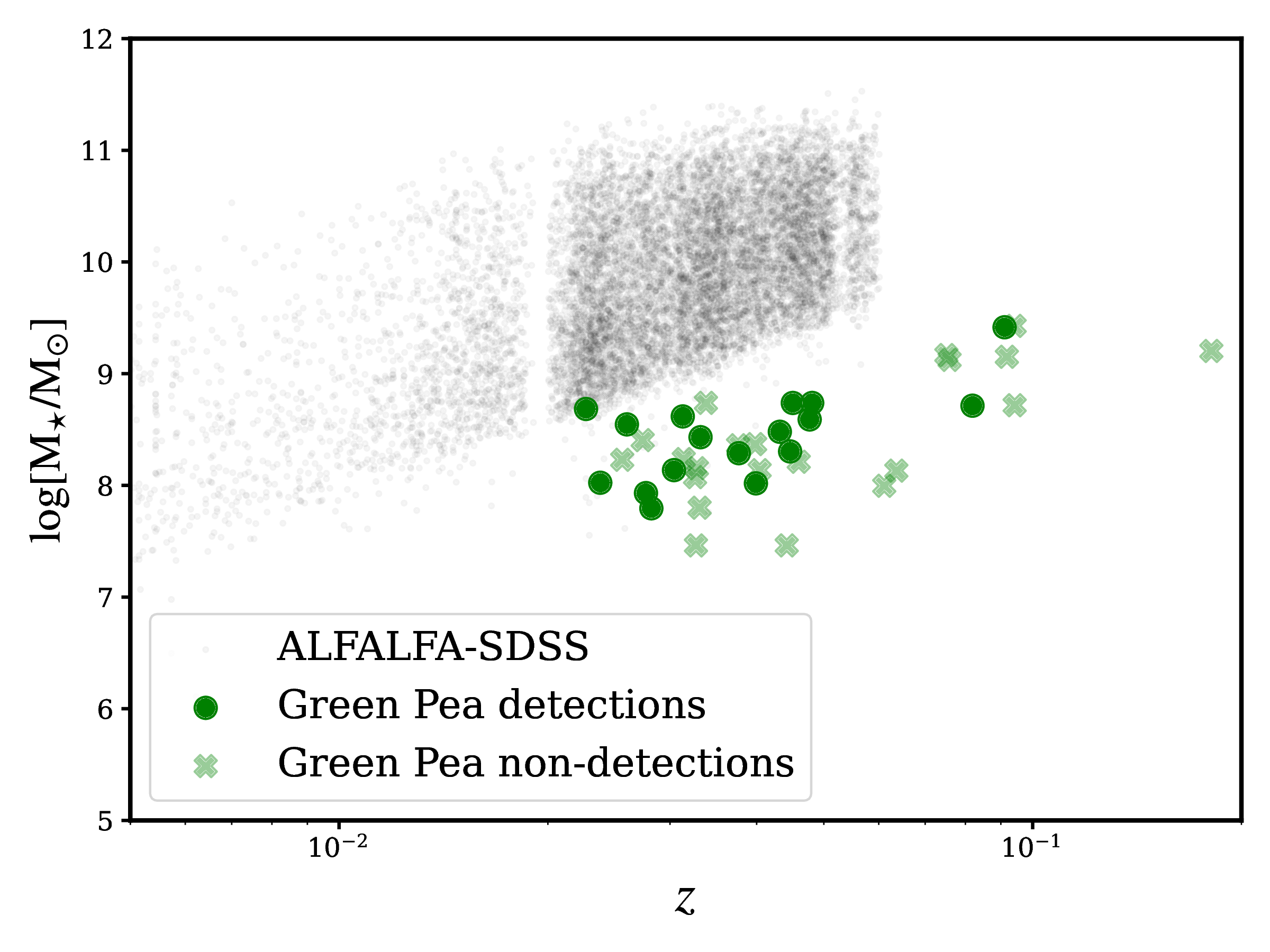}
      \caption{$z$ vs. stellar mass of the Green Pea galaxies and the comparison samples.
      The Green Pea galaxies with H{\sc i} 21 cm detections are marked with green circles, while the non-detections are marked with green crosses. 
      The ALFALFA-SDSS samples \citep{2020AJ....160..271D} are plotted with gray stars in the background.}
         \label{fig:z_mass}
  \end{figure}

For the comparison sample, we use the detections in the ALFALFA-SDSS catalog \citep{2020AJ....160..271D}.  
We adopt the stellar mass measurement using infrared photometry from the unWISE catalog based on Section 3.1 in \citet{2020AJ....160..271D} referred to as $M_{\star, \rm McGaugh}$ \citep{2015ApJ...802...18M}.
Our analysis employs the SFR measurements based on UV and IR photometry from \citet{2020AJ....160..271D} Section 3.2 quoted as $\rm SFR_{NUV_{cor}}$.
The mass and redshift distribution for all of these sources is demonstrated in Fig.\ref{fig:z_mass}.

\section{The scaling relations of the H{\sc i} gas fraction}
\label{sec:H I}
In this section, we revisit the scaling relation of the NUV-$r$ color with the H{\sc i} gas fraction and demonstrate the deviation of the observed H{\sc i} gas fraction of Green Pea galaxies to the predicted values.  
Then, we consider other factors that might impact the scaling relations of the H{\sc i} gas fractions and provide the scaling relations with linear combination forms, which combine color, surface mass density, surface brightness, or sSFR, that better predict the H{\sc i} gas fraction.

\subsection{The scaling relation of the H{\sc i} gas fraction with NUV-$r$ color}

Traditionally, different colors have been used to infer the H{\sc i} gas fraction of the star-forming galaxies, like the optical-optical color $u-r$ and the optical-NIR color $u-K$ in \citet{2004ApJ...611L..89K}, with a scatter of 0.4 dex. 
\citet{2005ApJ...619L...1M} demonstrate the correlation of the $g-i$ color and the NUV-$r$ color to the gas fraction in their Fig.5.  Later studies aim to improve the photometric estimators of H{\sc i} gas fraction by introducing the surface mass density and surface brightness.
\citet{2009MNRAS.397.1243Z} provide an estimator based on the optical color $g-r$ and the $i-$band surface brightness $\mu_{i}$ reaching a smaller scatter as 0.3 dex.
In \citet{2021A&A...648A..25Z}, both the $g-i$ color and the NUV-$r$ color are correlated with the H{\sc i} gas fraction, where a bluer color represents a higher gas fraction based on the low-$z$ calibrating sample consisting of 660 local galaxies.  
NUV-$r$ color is better to predict the H{\sc i} gas fraction with a larger dynamic span (see Fig.5 in \citet{2021A&A...648A..25Z}).  
\citet{2021A&A...648A..25Z} explain with the following statement: ``the ultra-violet emission of a galaxy mainly traces the light from the young stars in an optical thin environment and the extended gas in the outer disk of the galaxies are not absorbed by dust."

\begin{figure}[!ht]
  \centering
  \includegraphics[width=\hsize]{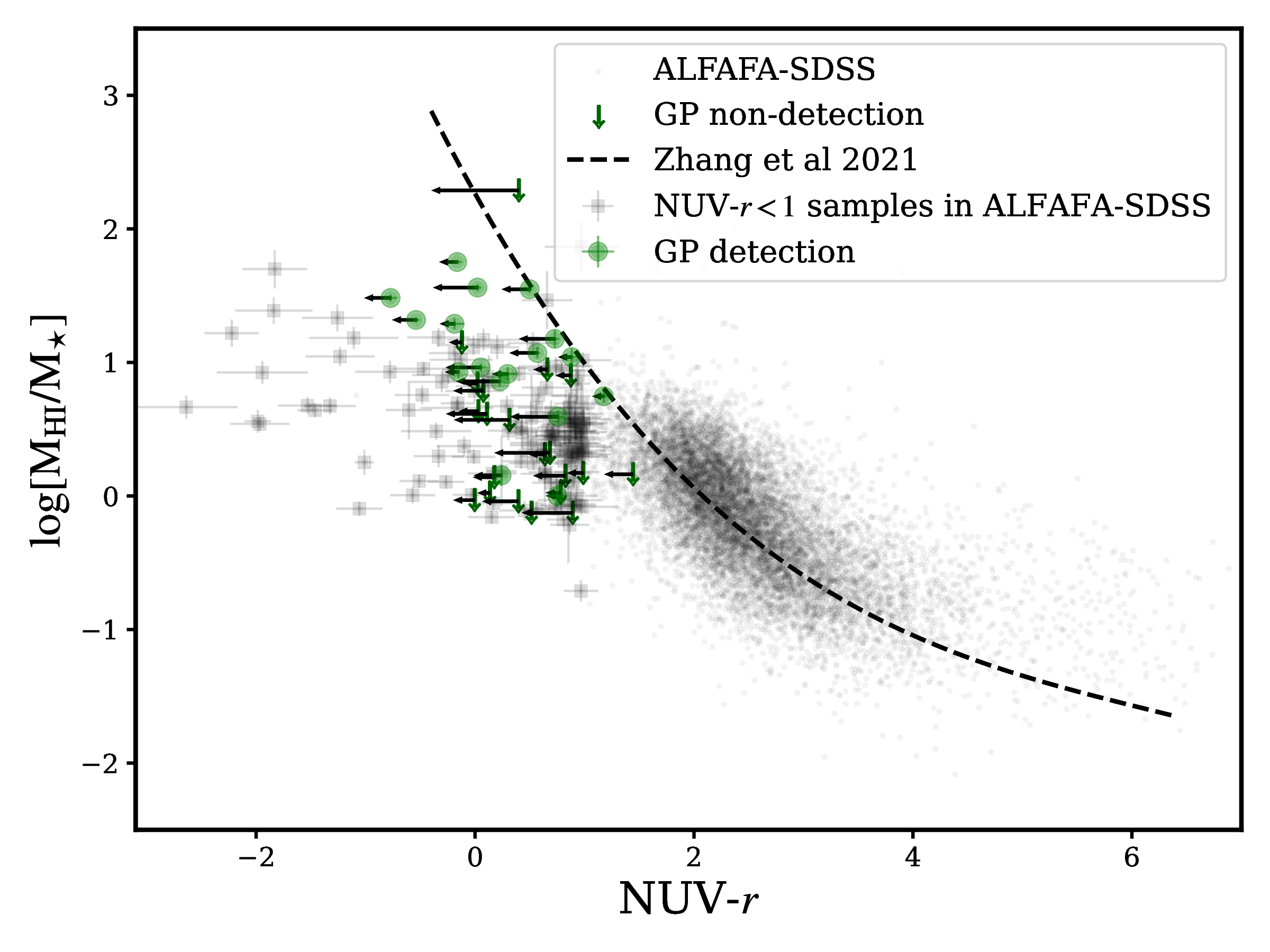}
      \caption{NUV-$r$ color vs the H{\sc i} gas fraction. 
      The dashed line shows the polynomial form NUV-$r$ color scaling relation from \citet{2021A&A...648A..25Z}.
      The Green Pea galaxies with H{\sc i} 21 cm emission lines detected are marked with green circles, while the non-detections are marked with green downward arrows. 
      To note the impact of the emission lines in the $r$-band photometry, we add the black arrows to show the NUV-$r$ color with the emission lines masked.
      Samples from the ALFALFA-SDSS catalog \citep{2020AJ....160..271D} are illustrated in the background with gray stars.
      The blue samples (NUV-$r$<1) from ALFALFA-SDSS are marked with gray squares, where the predicted $f_{\rm H\sc I}$ is higher than the observed H{\sc i} gas fraction.
      }
         \label{fig:NUV_r_H I_frac}
  \end{figure}

We apply the scaling relation of the NUV-$r$ color with the H{\sc i} gas fraction in Fig.\ref{fig:NUV_r_H I_frac}.  
It is noticed that the observed H{\sc i} gas fractions of the Green Pea galaxies deviate from the polynomial fit in \citet{2021A&A...648A..25Z}, where more H{\sc i} gas is predicted from the NUV-$r$ color. 

The emission lines contribute significantly to the broad-band color, therefore, we also display the NUV-$r$ color with the emission lines masked.  
The synthetic photometry of the $ugriz$ bands is obtained by convolving the spectra with the corresponding filters, and the magnitude difference is the difference between the synthetic photometry with the original spectra and the spectra with the emission lines masked.  
It is obvious in Fig.\ref{fig:NUV_r_H I_frac}, with the emission lines removed, the NUV-$r$ colors become bluer, and therefore deviate more from the \citet{2021A&A...648A..25Z} NUV-$r$ color scaling relation. 

Furthermore, there are a non-negligible amount (141) of blue samples (NUV-$r$<1) in the ALFALFA-SDSS catalog. These sources are marked with gray squares in Fig.\ref{fig:NUV_r_H I_frac}.
where the predicted $f_{\rm H\sc I}$ is higher than the observed H{\sc i} gas fraction for these blue samples. 
We have carefully checked these sources in the SDSS images and have found that the majority of the sources are more likely diffuse H {\sc ii} regions or bright star-forming regions in large galaxies. 
It is obvious in Fig.\ref{fig:NUV_r_H I_frac} that these samples also deviate from the polynomial form NUV-$r$ scaling relation of H{\sc i} gas fraction in \citet{2021A&A...648A..25Z}, where the observed values are lower than the predicted values, similar to the Green Pea galaxies.
With the optical emission lines removed, the NUV-$r$ color of these blue samples will become bluer as well, moving left in Fig.\ref{fig:NUV_r_H I_frac}.

\subsection{Other factors that impact the H{\sc i} gas fraction}
Due to the non-negligible offset between the observed H{\sc i} gas fraction and the predicted H{\sc i} from the NUV-$r$ scaling relation, we consider additional parameters to better estimate the H{\sc i} gas fraction for the Green Pea galaxies and the blue samples from ALFALFA-SDSS. 

\begin{figure}[!ht]
  \centering
  \includegraphics[width=0.8\hsize]{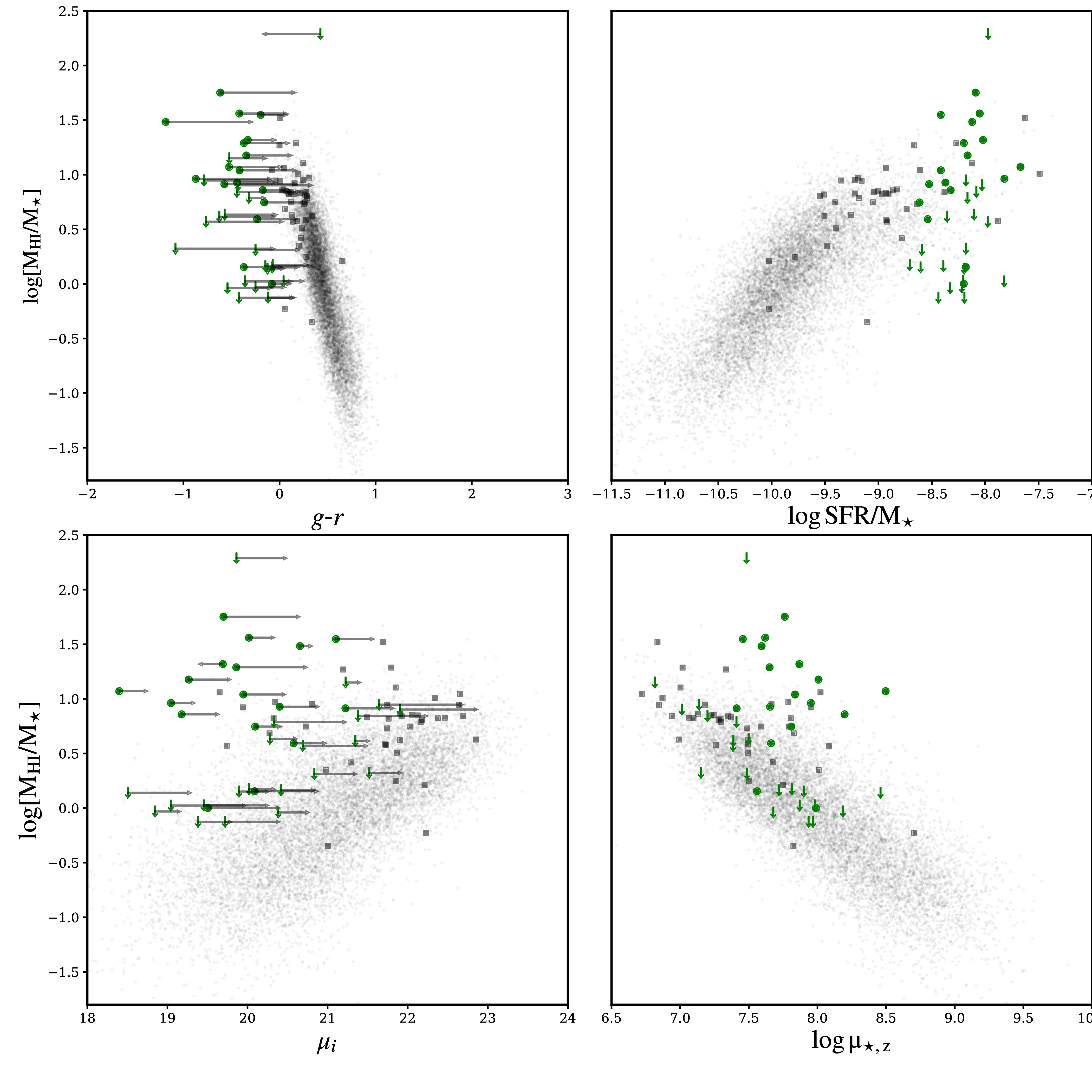}
  \caption{The distribution of the H{\sc i} gas fraction and the $g-r$ color (top left), the sSFR (top right), surface brightness (bottom left), and the surface mass density (bottom right).
  The sources with H{\sc i} 21 cm emission lines detected are marked with green circles, while the non-detections are marked with green downward arrows. 
    We also demonstrate $g-r$ color and the surface brightness $\rm \mu_i$ with the emission lines removed with the black arrows.
  The ALFALFA-SDSS samples are plotted with gray stars in the background, whereas the blue samples (NUV-$r$<1) are marked with gray squares.
  }
         \label{fig:comparison}
  \end{figure}

Four possible factors that will impact the measured NUV-$r$ with H{\sc i} gas fraction scaling relation are investigated: the $g-r$ color, the sSFR, surface brightness $\mu_i$ and the surface mass density $\log \mu_{\star}$ as discussed in \citet{2009MNRAS.397.1243Z,2010MNRAS.403..683C,2012MNRAS.424.1471L,2021A&A...648A..25Z}.  
Following the definitions in \citet{2009MNRAS.397.1243Z}, we define the surface mass density as $\rm \log(\mu_{\star,z}) = \log (\rm M_{\star}) - \log (2\pi \rm R^2_{50,z})$, where the $\rm M_{\star}$ is the stellar mass and the $\rm R_{50,z}$ is the radius (in units of kpc) enclosing 50\% of the total Petrosian $z-$band flux obtained from \texttt{photoObjAll} in SDSS.
Similarly, the surface brightness is defined as $\rm \mu_i = m_i + 2.5(2\pi \rm R^2_{50,i})$, where the $m_i$ is the apparent \texttt{cModelMag} of the $i-$band and the $\rm R_{50,i}$ is the radius (in units of kpc) enclosing 50 percent of the total Petrosian $z-$band flux in units of arcsecond. 
Green Pea galaxies are compact galaxies and are barely resolved in the SDSS images, where it is still reliable to use the measurement from SDSS.
Three Green Pea galaxies have obtained acquisition images from the COS instrument in the UV of \textit{HST} \citep{2019hst..prop16054K}, with 200 s exposure time each. 
We use SExtractor \citep{1996A&AS..117..393B} to measure the Petrosian $R_{50,\rm HST}$ for these three sources: J1108+2238 is 0.964 arcsec, J1345+0442 is 0.418 arcsec, and J1455+3808 is 0.577 arcsec.
These radii in the UV $R_{50,\rm HST}$ are smaller than the $R_{50,z}$.
The comparison of the H{\sc i} gas fraction with the $g-r$ color, the sSFR, surface mass density, and the surface brightness for Green Pea galaxies and the ALFALFA-SDSS samples are demonstrated in Fig.\ref{fig:comparison},
as well as the distribution of the $g-r$ color and the surface brightness with the emission lines removed for the Green Pea galaxies.
With the emission lines removed, the Green Pea galaxies occupy similar $g-r$ color, which intrinsically traces mainly the small- to intermediate-mass stars, similar to the ALFALFA-SDSS samples.  However, the Green Pea galaxies are dominated by massive stars from ongoing starbursts, thus they deviate from the ALFALFA-SDSS samples in the NUV-$r$ color with the emission lines removed. 

\begin{figure}[!ht]
  \centering
  \includegraphics[width=\hsize]{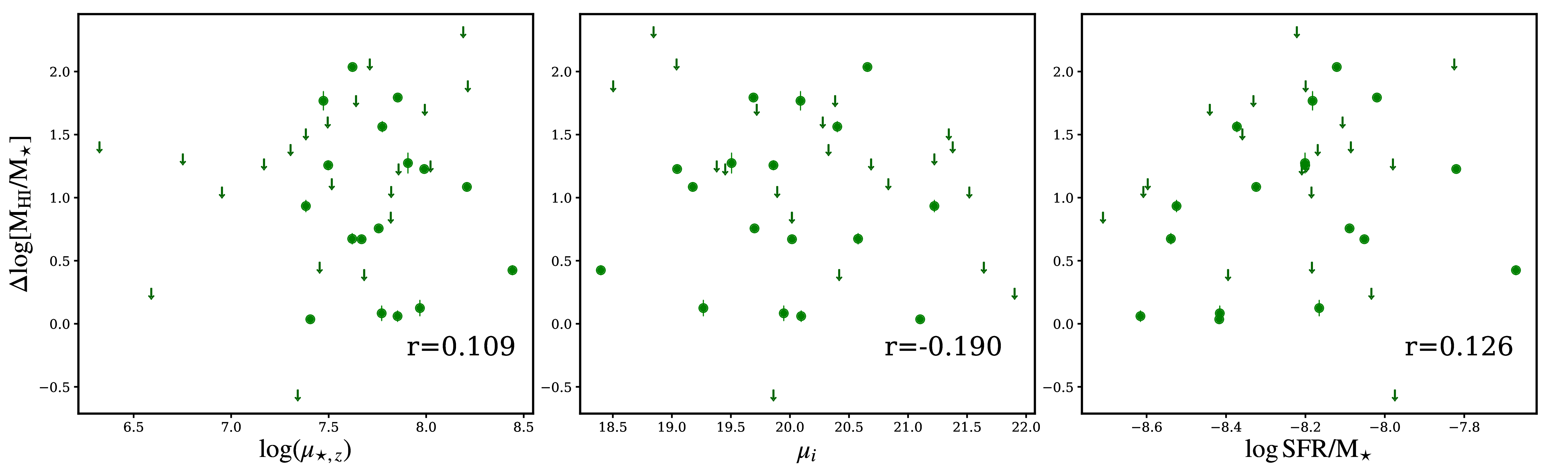}
  \caption{The relation of the H{\sc i} gas fraction offset (prediction - observed) with the surface mass density (left), the surface brightness (middle), and the sSFR (right).  
  The sources with H{\sc i} 21 cm emission lines detected are marked with green circles, while the non-detections are marked with green downward arrows. 
  We also mark Pearson's correlation coefficients in each subplot.
  }
         \label{fig:offset}
  \end{figure}

It is natural to consider including the surface mass density, the surface brightness, and the sSFR into the scaling relation of the H{\sc i} gas fraction.
We consider the differences between the H{\sc i} gas fraction predicted by the NUV-$r$ color and the observed gas fraction as the offset and demonstrate the relation of these three quantities with the offset in Fig.\ref{fig:offset}.
  There is a positive correlation between the offset and the surface mass density and the sSFR as demonstrated in the left and right panels of Fig.\ref{fig:offset}, while the surface brightness anti-correlates with the surface brightness in the middle panel. 

\subsection{The scaling relations of the H{\sc i} gas fraction with linear combination forms}
\begin{figure*}[!ht]
  \centering
  \includegraphics[width=\hsize]{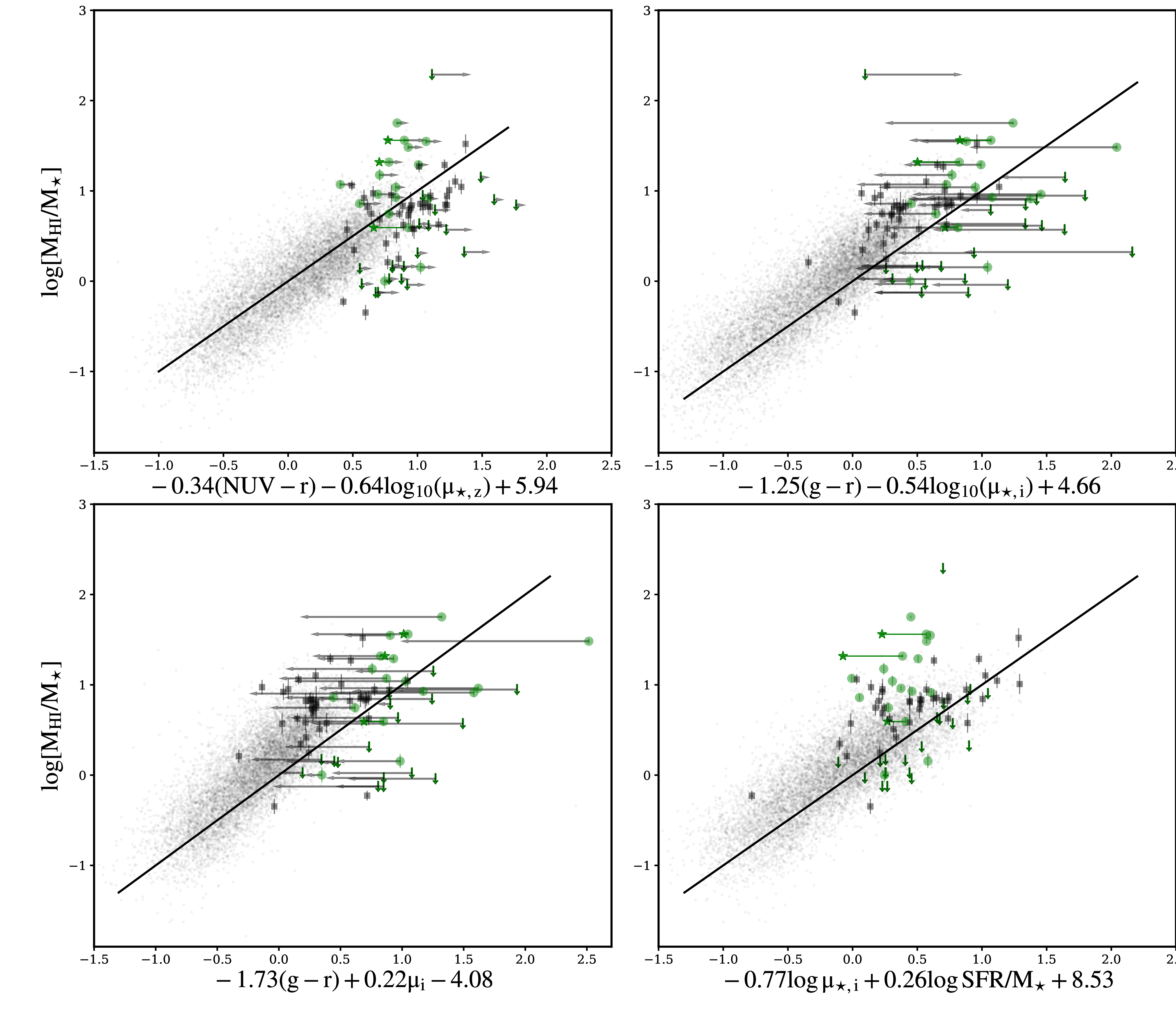}
  \caption{The relation between the H{\sc i} gas fraction and the scaling relations linearly combining different physical quantities. 
  From top left to bottom right: (1) the NUV-$r$ color and stellar surface mass density in the $z$-band, (2) $g-r$ color and the stellar surface mass density, (3) $g-r$ color and the $i-$band surface brightness, and (4) sSFR and stellar surface mass density.
  For the three sources with morphological information from \textit{HST}, we calculate the surface mass density and surface brightness with the $R_{50,\it HST}$ and mark them with green stars connecting the estimator with $R_{50}$ from SDSS.
  To check the impact of the emission lines on the optical broad-band photometry, we also demonstrate the estimators with the emission lines removed, which is marked by the black solid arrows.
  The scaling relations are marked with black solid lines in each subplot.  
  The ALFALFA-SDSS samples are plotted with gray stars in the background, whereas the blue samples (NUV-$r$<1) are marked with gray squares.
  }
         \label{fig:color_surface_mass_density}
  \end{figure*}

We test four sets of linear combination H{\sc i} gas fraction estimators based on the relations in \citet{2009MNRAS.397.1243Z,2021A&A...648A..25Z}.
To be noted, that there are other forms of linear combination estimators in \citet{2010MNRAS.403..683C,2012MNRAS.424.1471L}.
  Following the relation in \citet{2021A&A...648A..25Z}, for the blue sources with NUV-$r<$3.5, we check whether a linear combination of NUV-$r$ color and the stellar surface mass density will predict the H{\sc i} gas fraction more accurately. 
  The top left panel in Fig.\ref{fig:color_surface_mass_density} demonstrates the distribution of the Green Pea galaxies and the blue samples (NUV-$r<3.5$) from ALFALFA-SDSS catalog \citep{2020AJ....160..271D}, with the form of $\rm -0.34(NUV-r) - 0.64 \log(\mu_{\star,z}) + 5.94 $.  
  
  Three other sets of the linear combination H{\sc i} gas fraction estimators from \citet{2009MNRAS.397.1243Z} that use the optical color, sSFR, surface brightness, and surface mass density are also applied for the Green Pea galaxies.
  The Green Pea galaxies with H{\sc i} 21 cm detections and non-detections distribute above and below the linear relation, in agreement with the linear estimator prediction.
  Among the four sets of linear combination gas fraction estimators, the first estimator $\rm -0.34(NUV-r) - 0.64 \log(\mu_{\star,z}) + 5.94 $ \citep{2021A&A...648A..25Z}, and the fourth estimator $\rm -0.77 \log \mu_{\star,i} + 0.26 \log SFR/M_{\star}+8.53$ \citep{2009MNRAS.397.1243Z} result in smaller scatters.
    Even with the impact of the emission lines taken into consideration, the first estimator can still predict the gas fraction robustly. 
  This is reasonable due to (1) a broader dynamic range of the NUV-$r$ color compared with the $g-r$ color \citep{2021A&A...648A..25Z}, (2) the NUV-$r$ color is dominated by the emission from massive stars compared with $g-r$ where mainly from the low- to intermediate-mass (discussed in Section 3.2), (3) this is a more dedicated scaling relation tuned towards the blue samples (NUV-$r$ <1). 

However, these measurements are based on barely resolved \footnote{We obtain the FWHM of the PSF for the same band from \texttt{photoObjAll} in SDSS and find the $R_{50,z}$ of the Green Pea galaxies are larger than half of the \texttt{psfFWHM}.} measurements of these compact galaxies, where the measured $R_{50}$ is not accurate. 
 For three Green Pea galaxies with morphological information from \textit{HST}, we calculate the surface mass density and surface brightness with the $R_{50,\it HST}$.  
 In Fig.\ref{fig:color_surface_mass_density}, these measurements are marked with green stars, they are connected to the measurements based on SDSS $R_{50}$ with green lines.
More accurate optical imaging with the upcoming ground-based and space-based facilities will be helpful to accurately measure the surface mass density and the surface brightness of the Green Pea galaxies.
This will be helpful for better calibrating the scaling relations of the H{\sc i} gas fraction and reducing the scatter, serving as cheap and convenient $f_{\rm H\sc I}$ estimators for potential application to large samples of star-forming galaxies.

\section{Result}
\label{sec:result}
 Based on the archival data of the H{\sc i} mass measurements from \citet{2021ApJ...913L..15K}, we compare 38 Green Pea galaxies, including 17 detections and 21 non-detections, to the comparison sample of local star-forming galaxies 
 from ALFALFA-SDSS \citep{2020AJ....160..271D}, to check the H{\sc i} gas fraction scaling relations.
 We find that the Green Pea galaxies deviate from the \citet{2021A&A...648A..25Z}  NUV-$r$ polynomial form, where the observed gas fractions are lower than the predictions, even with the emission lines removed from the $r-$band photometry. 

 The offsets between the predicted H{\sc i} gas fraction and the measured H{\sc i} gas fraction correlate with the surface mass density, the surface brightness, and the sSFR.
 Therefore, these three quantities are included in the scaling relations of the H{\sc i} gas fraction with linear combination forms.
 The forms of $\rm -0.34(NUV-r) - 0.64 \log(\mu_{\star,z}) + 5.94 $, and $\rm -0.77 \log \mu_{\star,i} + 0.26 \log SFR/M_{\star}+8.53$ \citep{2009MNRAS.397.1243Z,2021A&A...648A..25Z}, better predict the H{\sc i} gas fraction of the Green Pea galaxies. 
In order to obtain accurate linear combined forms with smaller scatter, higher-resolution photometry from space-based telescopes is needed.

There are a large amount of optically identified Green Pea galaxies \citep{2019ApJ...872..145J,2022ApJ...927...57L}, where the median NUV-$r$ color for these samples is 1.13 and presumably high H{\sc i} gas fractions.
We have verified in this work that the predicted H{\sc i} gas fractions from the scaling relations with the linear combination forms agree well with the observed H{\sc i} gas fraction.
The H{\sc i} gas fraction of these Green Pea galaxies and other types of blue star-forming galaxies can be estimated and applied for other analyses. 

\begin{acknowledgements}
This work is supported by the National Science Foundation of China (Nos. 12273075 and 12090041). 
This publication makes use of data products from the \textit{Wide-field Infrared Survey Explorer}, which is a joint project of the University of California, Los Angeles, and the Jet Propulsion Laboratory/California Institute of Technology, funded by the National Aeronautics and Space Administration.
W. Zhang acknowledges support from the National Key R\&D Program of China (No. 2021YFA1600401, 2021YFA1600400), and the Guangxi Natural Science Foundation (No. 2019GXNSFFA245008).
This research uses \texttt{Astropy} 
\citep{2013A&A...558A..33A,2018AJ....156..123A}, \texttt{TOPCAT} \citep{2005ASPC..347...29T}, \texttt{Scipy} \citep{2020NatMe..17..261V}, \texttt{Numpy} \citep{2020Natur.585..357H}, \texttt{lifelines} \citep{Davidson-Pilon2019},
\texttt{pandas} \citep{mckinney-proc-scipy-2010,reback2020pandas} and \texttt{Matplotlib} \citep{Hunter:2007}
\end{acknowledgements}

\bibliographystyle{aasjournal}
\bibliography{GP_new}

\newpage

\label{lastpage}

\end{document}